\documentclass[graybox]{svmult}

\usepackage{mathptmx}       % selects Times Roman as basic font
\usepackage{helvet}         % selects Helvetica as sans-serif font
\usepackage{courier}        % selects Courier as typewriter font
\usepackage{type1cm}        % activate if the above 3 fonts are
\usepackage{makeidx}         % allows index generation
\usepackage{graphicx}        % standard LaTeX graphics tool
\usepackage{multicol}        % used for the two-column index
\usepackage[bottom]{footmisc}% places footnotes at page bottom

\usepackage{cite}

\usepackage{graphicx}
\usepackage{color}
\usepackage{array}
\usepackage{amsmath}
\usepackage{amssymb,latexsym}
\usepackage{mathrsfs}
\usepackage{cancel}
\usepackage{url}
\def \CSmooth(#1,#2){\mathcal{C}_{#1,#2}}

\def \Mgale(#1,#2){M_{#1}^{#2}}
\def \Ngale(#1,#2){\mathcal{N}_{#1}^{#2}}

\newcommand{\domH}{C^1([0,T],{\rm dom}\Delta)}

\makeindex             % used for the subject index
\begin{document}

\title*{From non-symmetric particle systems to non-linear PDEs on fractals}
\author{Joe P. Chen, Michael Hinz, Alexander Teplyaev}

\institute{Joe P. Chen: Department of Mathematics, Colgate University, Hamilton, NY 13346, USA, \email{jpchen@colgate.edu}
\\  Michael Hinz: Fakult\"at f\"ur Mathematik, Universit\"at Bielefeld, Postfach 100131, 33501 Bielefeld, Germany, \email{mhinz@math.uni-bielefeld.de}
\\ Alexander Teplyaev: Department of Mathematics, University of Connecticut, Storrs, CT 06269, USA, \email{teplyaev@uconn.edu}
\\Research supported in part by NSF grants 1262929, 1613025. 
}

\maketitle\abstract{
 	We present new results and challenges in obtaining hydrodynamic limits for non-symmetric (weakly asymmetric) particle systems (exclusion processes on pre-fractal graphs) converging to a non-linear heat equation.   
 	We discuss a joint density-current law of large numbers and a corresponding large deviation principle.   }
%\section{Introduction}
%
%Interacting particle systems 
%Mathematical challenges. On $SG$

%\section{Setup and main results}

%\subsection

\bigskip\noindent{\bf Exclusion process on a weighted graph.}
We consider a locally finite connected (simple and undirected) graph $\Gamma= (V, E)$ with vertex set $V$ and edge set $E$ and endowed with conductances ${\bf c} = (c_{xy})_{xy\in E}$ satisfying $c_{xy}>0$. 
The pair $(\Gamma, {\bf c})$ is called a \emph{weighted graph}. Suppose that $H: [0,T]\times V \to \mathbb{R}$ is a given   function with the abbreviated notation $H_t := H(t,\cdot)$. The \emph{weakly asymmetric exclusion process} (WASEP) on $(\Gamma, {\bf c})$ associated with $H$ is the Markov chain $(\eta_t)_{t\geq 0}$ on $\{0,1\}^{V}$ with time-dependent generator $\mathcal{L}_{(\Gamma, {\bf c}),H_t}^{\rm EX}$ defined on functions $f: \{0,1\}^{V} \to \mathbb{R}$ by
\[
\left(\mathcal{L}_{(\Gamma, {\bf c}),H_t}^{\rm EX} f\right)(\eta) = \sum_{xy\in E} c_{xy} \psi_{xy}(H_t,\eta) [f(\eta^{xy})-f(\eta)],
\]
where $\psi_{xy}(H_t,\eta) =  \exp\left\{\left(\eta(y)-\eta(x)\right) \left(H_t(x)-H_t(y)\right)\right\}$ and
\[
\eta^{xy}(z) = \left\{\begin{array}{ll}\eta(y),& \text{if}~z=x, \\ \eta(x), & \text{if}~z=y, \\ \eta(z), & \text{otherwise}.\end{array} \right.
\]
We can think of $\eta(z)$ as the occupation variable which returns $1$ (resp.\@ $0$) when $z$ is occupied with a particle (resp.\@ empty). The configuration $\eta^{xy}$ is obtained by exchanging the occupation variables $\eta(x)$ and $\eta(y)$ in $\eta$. At time $t$ such a transition occurs with rate $c_{xy}\psi_{xy}(H_t,\eta)$, where $\psi_{xy}$ encodes the (weak) asymmetry between the hopping rates from $x$ to $y$ and from $y$ to $x$. When $H\equiv 0$ we obtain the \emph{symmetric exclusion process} (SEP) on $(\Gamma,{\bf c})$.

We also define the \emph{boundary-driven} exclusion process. Declare a nonempty subset $\partial V \subset V$ to be the boundary set. Given the aforementioned exclusion process, we add a birth-and-death process to each boundary point $a\in \partial V$; that is, we consider the Markov chain on $\{0,1\}^{V}$ generated by
\[
\mathcal{L}^{\rm bEX}_{(\Gamma,{\bf c}),H_t} = \mathcal{L}^{\rm EX}_{(\Gamma,{\bf c}),H_t} + \mathcal{L}^{\rm b}_{\partial V}
,\]
where for any function $f: \{0,1\}^{V} \to\mathbb{R}$,
\[
(\mathcal{L}^b_{\partial V} f)(\eta)  = \sum_{a\in \partial V} [\lambda_-(a) \eta(a) + \lambda_+(a) (1-\eta(a))] [ f(\eta^a)-f(\eta)], 
\]
with $\lambda_+(a)>0$ (resp.\@ $\lambda_-(a)>0$) representing the birth (resp.\@ death) rate at $a$, and
\[
\eta^a(z) = \left\{\begin{array}{ll} 1-\eta(a),& \text{if}~z=a,\\ \eta(z), & \text{otherwise.}\end{array}\right.
\]
%In what follows, w
We assume 
%all conductances to equal one, $c_{xy}=1$, and drop the explicit mention of ${\bf c}$. We also assume 
that the relative boundary transition rates are bounded away from $0$ and $\infty$, i.e. we assume there exists $\gamma\in [1,\infty)$ such that $\gamma^{-1}\leq \frac{\lambda_+(a)}{\lambda_-(a)} \leq \gamma$ for all $a\in \partial V$.

%There are two key properties of the SEP. First, the total number of particles is conserved in the process. Second, the process is reversible with respect to any constant-density product Bernoulli measure $\nu_\alpha$ on $\{0,1\}^{V(\Gamma)}$, $\alpha\in [0,1]$, which has marginal $\nu_\alpha\{\zeta: \zeta(x)=1\}=\alpha
%$ for all $x\in V(\Gamma)$. 

%We introduce the Dirichlet energy
%\begin{align}
%\mathcal{E}^{\rm EX}_{(G,{\bf c}), \nu_\alpha}(f) = \nu_\alpha\left[f(-\mathcal{L}^{\rm EX}_{(G,{\bf c})} f) \right] = \frac{1}{2}\sum_{xy\in E} c_{xy} \,\nu_\alpha\left[(\nabla_{xy} f)^2\right], \quad f: \{0,1\}^V \to \mathbb{R},
%\end{align}

%\subsection

\begin{figure}[h]
	\centering
	\includegraphics[width=0.4\textwidth]{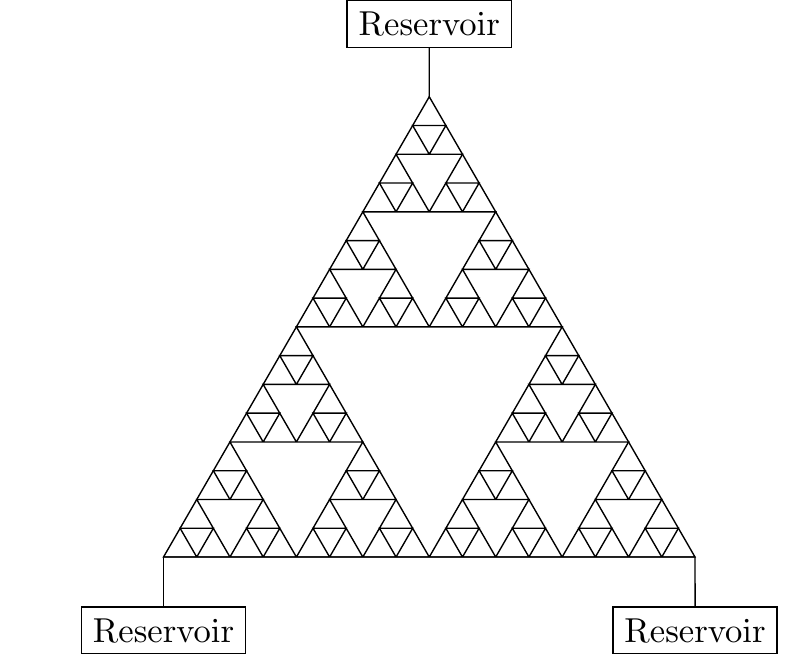}
	\caption{The pre-Sierpinski gasket graph $\Gamma_4$, indicating the three reservoirs which fix the particle densities at the boundary points $\{a_0,a_1,a_2\}$.}
	\label{fig:SGreservoir}
\end{figure}

\noindent{\bf Analysis on the Sierpinski gasket and exclusion processes on pre-fractal graphs.}
We now turn to the Sierpinski gasket ($SG$). Let $a_0$, $a_1$, $a_2$ be the vertices of an equilateral triangle in $\mathbb{R}^2$, and $\Gamma_0$ be the complete graph on the vertex set $V_0 = \{a_0, a_1, a_2\}$. Define the contracting similitude $\Psi_i : \mathbb{R}^2 \to\mathbb{R}^2$, $\Psi_i(x) = \frac{1}{2}(x-a_i) + a_i $ for each $i\in \{0,1,2\}$. For $N\geq 1$, we define $\Gamma_N = (V_N, E_N)$ inductively via the formula $\Gamma_N = \bigcup_{i=0}^2 \Psi_i(\Gamma_{N-1})$. The sequence $(\Gamma_N)_{N\geq 0}$ forms a graphical approximation of the $SG$ fractal $K$, which is the unique compact set satisfying $K= \cup_{i=0}^2 \Psi_i(K)$. By $\mu$ we denote the standard self-similar Borel probability measure on $K$.

A well-known result of Barlow-Perkins \cite{BarlowPerkins} states that if for each $N$ we write $(X_t^N)_{t\geq 0}$ to denote the natural symmetric random walk on the approximating graph $\Gamma_N$, then the sequence of processes $(X_{5^N \cdot}^N)_{N\geq 0}$ is tight and converges to a generic diffusion process (a ``Brownian motion'') $(X_t)_{t\geq 0}$ on $K$. The time acceleration factor $5^N$ can also be observed in the analytic approach of Kigami \cite{KigamiBook}, in the sense that the sequence of graph energies 
%$(\mathcal{E}_N)_{N\geq 0}$ with
$\mathcal{E}_N(f)=\frac{5^N}{3^N}\sum_{xy\in E_N} [f(x)-f(y)]^2$, $ f: V_N\to\mathbb{R}
$
is monotone and converges to a limit energy form $(\mathcal{E},\mathcal{F})$ on $K$. This form is a resistance form in the sense of Kigami, \cite{Kigamiresistance}, and in particular, it satisfies  
the Sobolev embedding $\mathcal{F} \subset C(K)$. Moreover, it defines a strongly local regular Dirichlet form on $L^2(K,\mu)$. This allows to define the standard (Dirichlet) Laplacian $\Delta$ by a Gauss-Green formula: given $f\in C(K)$, we say that $u\in {\rm dom}\:\Delta$ with $\Delta u=f$ if 
$\mathcal{E}(u,v) =-\int_K\, f v\,d\mu$ 
for all $v\in \mathcal{F}$ with $v|_{V_0}=0$. Then $(\Delta, {\rm dom}\:\Delta)$ is a non-positive self-adjoint operator on $L^2(K,\mu)$, and it is just the $L^2(X,\mu)$-infinitesimal generator of the diffusion $(X_t)_{t\geq 0}$. For more details see \cite{BarlowStFlour, KigamiBook,Kigamiresistance,StrichartzBook}. 

We also need the notions of gradient and divergence. On general weighted graphs 
%$(\Gamma,{\bf c})$ 
they are explained in \emph{e.g.\@} \cite{LyonsPeres}. We fix an orientation for each edge $xy\in E_N$ of the approximating graph $\Gamma_N=(V_N,E_N)$, and we may assume that the resulting oriented edge, denoted by $\overrightarrow{xy}$, has initial vertex $x$ and terminal vertex $y$. 
%{\color{blue} (Q from Michael: I suggest not to use the symbol $\overrightarrow{E}$. Is it really needed if orientations are fixed ?)} 
By $\ell^2_-(E_N)$ we denote the space of functions $\theta:E_N\to\mathbb{R}$ that are antisymmetric under a change of orientation, \emph{i.e.\@} satisfy $\theta(\overrightarrow{yx}) = -\theta(\overrightarrow{xy})$ for all $\overrightarrow{xy}\in E$. On each approximating graph $\Gamma_N$ we can define the \emph{discrete gradient} $\partial_N: \ell^2(V_N)\to\ell^2_-(E_N)$ 
%and the \emph{discrete divergence} $\partial_N^*: \ell^2_-(E_N) \to \ell^2(V_N)$ 
by
$
(\partial_N f)(\overrightarrow{xy}) = 
%5^{N/2}
\frac{5^N}{3^N}
[f(y)-f(x)]
%\quad 
%\text{ and }\quad (\partial_N^* {\bf F})(x) = 5^{N/2}\sum_{y \in V_N: \overrightarrow{xy}\in E_N} {\bf F}(\overrightarrow{xy})
$.
%, \]
%{\color{blue} (Q from Michael: Do you insist on this arrangement of prefactors ? Please also check that writing summation this way goes ok with the orientations / please change.)}
%where the factor $5^{N/2}$ reflects the diffusive scaling. It is direct to verify that $\partial_N$ and $-\partial_N^*$ are adjoints of each other, $\langle \partial_N f, {\bf F}\rangle_{\ell^2_-(E_N)} = -\langle f, \partial_N^* {\bf F}\rangle_{\ell^2(V_N)}$.
%{\color{blue} (Q from Michael: Please check sign.)}
%Moreover, if $\mu_N$ denotes the uniform probability measure on $E_N$, then 
%\[\|\partial_N f\|_{L^2(E_N,\mu_N)}^2 =\mathcal{E}_N(f).\]
%\begin{align}
%\langle \partial_N f_1, \partial_N f_2\rangle_{E_N} = \mathcal{E}_{G_N}(f_1,f_2), &\quad f_1, f_2 \in \Omega_0(G_N).\\
%\langle  \partial_N f, {\bf F}\rangle_{E_N} = -5^{N/2} \sum_{x\in V_N} f(x) \sum_{y \simN x} {\bf F}(\overrightarrow{xy})  = -\langle f, \partial_N^* {\bf F}\rangle_{V_N}, & \quad f\in \Omega_0(G_N), \quad {\bf F}\in \Omega_1(G_N).
%\end{align}
The continuum analogs of $\partial_N$ 
%and $\partial_N^*$ 
on the limiting fractal $K$ can be introduced and studied via Dirichlet form theory, \emph{cf.\@} \cite{x,CipSau03, CGIS13, CGIS14, IRT12, HinzRocknerTeplyaev, HinzTeplyaevNS, HinzTeplyaevReview, HinzKelleherTeplyaev}. There exist a certain \emph{Hilbert bimodule $\mathcal{H}$ of generalized $L^2$-vector fields associated with $(\mathcal{E}, \mathcal{F})$} and a derivation $\partial: \mathcal{F} \to \mathcal{H}$ that satisfies 
$
\|\partial f\|_{\mathcal{H}}^2=\mathcal{E}(f)$, $ f\in\mathcal{F}$.
The adjoint $-\partial^*: \mathcal{H}\to \mathcal{F}$ is then defined in the usual way by duality. In particular,  
\[
\langle \partial f,g \partial h\rangle_\mathcal{H} = -\langle f, \partial^*(g\partial h)\rangle_{L^2(K,\mu)}= \int_K \, g d\Gamma(f,h)
\]
for all $f,g,h\in \mathcal{F}$, where $\Gamma(f,h)$ denote the \emph{mutual energy measure} of $f$ and $h$, \cite{ChenFukushima,MR,StrichartzBook}. The operators $\partial$ and $-\partial^*$ may be viewed as (abstract) \emph{gradient} and \emph{divergence operators} on $K$. 
By construction we have 
$\lim_{N\to\infty}\|\partial_N f\|
%_{L^2(E_N,m_N)} 
= \|\partial f\|_\mathcal{H}$ with an appropriate definition of norm involved, 
However, unlike in the Euclidean setting, we do not have the pointwise convergence $\partial_N f\to\partial f$.

We consider the boundary-driven WASEP $(\eta_t^N)_{t\in[0,T]}$ on $\Gamma_N$ generated by the operator $5^N\left(\mathcal{L}^{\rm EX}_{\Gamma_N, H_t} + \mathcal{L}^{b}_{V_0} \right)$.
assuming that $H\in C^1([0,T],{\rm dom}\Delta)$,
%Here $H$ is drawn from the space
%\begin{multline}\label{eq:DH}
%\mathcal{D}_H :=\bigl\{ H:[0,T]\times K\to\mathbb{R}: \text{for each $x\in K$, $H_\cdot(x)$ is strongly differentiable,}\\ 
%\text{$H_t \in {\rm dom}\:\Delta$ for all $t\in[0,T]$, and $\esssup_{t\in(0,T)} \mathcal{E}(H_t) <\infty$}\bigr\} ,
%\end{multline}
and that the boundary set $V_0$ and the rates $\{\lambda_\pm(a_i): i\in\{0,1,2\}\}$ are fixed for all $N$. See Figure \ref{fig:SGreservoir} for a schematic picture.

We are interested in two 
observables (empirical measures)
%(measures on $K $) 
in the exclusion process: the \emph{empirical density}
and the \emph{empirical integrated current}
\[
\pi^N_t(A) = \frac{1}{|V_N|}\sum_{x\in V_N} \eta^N_t(x) \mathbf{1}_A(x), \quad A\subset K, 
\]
 \[
{\bf W}_t^N (B)= \frac{1}{%5^{N/2}
	|E_N|} \sum_{\overrightarrow{xy}\in E_N} W^N_t(\overrightarrow{xy}) \mathbf{1}_B(\overrightarrow{xy}) 
,
\]
where $B$ is a subset of oriented edges. Here the \emph{current} $W^N_t(\overrightarrow{xy})$, for each oriented edge $\overrightarrow{xy}$ in $E_N$, is the net number of particle jumps along $\overrightarrow{xy}$ (\emph{i.e.\@} \#(jumps from $x$ to $y$) $-$ \#(jumps from $y$ to $x$)) in the time interval $[0,t]$. 
We 
%then define the 
%taking into account the scaling factor $5^{-N/2}$. {\color{blue} (Q by Michael: What is $B$ ? A given finite set of edges ? Is the empirical integrates current sth like a measure on the set of edges ?)}
% and 
note the mass conservation law %{\color{blue} [check the statement at the boundary]}
$
\sum_{y\in V_N: \overrightarrow{xy}\in E_N} W^N_t(\overrightarrow{xy}) = -\left[\eta^N_t(x) -\eta^N_0(x)\right] 
$ for $x\in V_N\backslash V_0$.
%{\color{blue} Q by Michael: Please check / correct again the notation using oriented edges.)}

%\subsection

\bigskip\noindent{\bf Hydrodynamic limit.}
The density-current pair $\left(\pi^N_t,{\bf W}^N_t\right)$ satisfies a law of large numbers. 
%Loosely speaking, w
We fix a macroscopic density $\rho_0$ %{\color{blue} (Q by Michael: If the solution is denoted by $\varrho$ this clashes with the notation $\rho_0=\rho(0,\cdot)$. That below we require the solution to equal $\rho_0$ at zero, should be visible as a condition. I suggest to write $\varrho^H$ for the solution already in the equation.)}
satisfying the boundary condition 
\[\rho_0(a_i) = \bar{\rho}_i := \frac{\lambda_+(a_i)}{\lambda_+(a_i)+\lambda_-(a_i)}\] 
on $V_0$, and assume that the sequence of initial densities $(\pi^N_0)_{N\geq 1}$ converges weakly to $\rho_0 \,d\mu$. We then claim that $(\pi^N_t)_{N\geq 1}$ converges to the weak solution $\rho^H$ of the nonlinear parabolic equation
\begin{align}
\label{eq:hydrodynamiceqn}
\left\{\begin{array}{lll} \partial_t\rho^H(t,x) &= \Delta\rho^H(t,x) - \partial^*\left(\chi(\rho^H(t,x)) \partial H(t,x)\right) & \text{on}~(0,T)\times (K\setminus V_0), \\
 \rho^H(0,x) &= \rho_0(x),& \text{on}~K\setminus V_0, \\ 
 \rho^H(t,a_i) &= \bar\rho_i& \text{on}~(0,T),\end{array}\right.
\end{align}
where $\chi:\mathbb{R}\to [0,1]$ is defined by $\chi(s):=(s(1-s))_+$ 
%{\color{blue} 
	%(Q by Michael: Please check that taking the positive part  is ok.)} 
The quantity $\chi(\rho^H)$ is the mobility of the exclusion process. In addition, the time derivative of ${\bf W}^N_t$ converges to the vector field ${\bf J}_t=-\partial\rho^H_t+\chi(\rho_t^H) \partial H_t$, which satisfies the macroscopic continuity equation $\partial_t \rho_t^H + \partial^* {\bf J}_t=0$. An important caveat is that these equations are only interpreted in the weak sense, not in the pointwise sense.

Turning to the formal description, we set 
\[C_e(K):= \{f\in C(K): \text{there exists $\varepsilon>0$ such that } \varepsilon\leq f\leq 1-\varepsilon\}.\] 
Fix $\rho_0 \in C_e(K)$ which satisfies the boundary condition $\rho_0(a_i) = \bar{\rho}_i$. Consider, for every $N$, the boundary-driven WASEP $(\eta^N_t)_{t\in [0,T]}$ whose initial configurations is $\eta^N$, 
%{(\color{blue} Q by Michael: Maybe better write $\eta_0^N$ here to be consistent with the formula for $\pi_t^N$ ?)}, 
and denote the corresponding law by $\mathbb{P}^N_{\eta^N}$. We assume that $(\eta^N)_{N\geq 0}$ is associated with $\rho_0$ in the sense of weak convergence, i.e. that 
\[\lim_{N\to\infty}\langle f,\pi^N_0\rangle = \langle f,\rho_0 \,d\mu\rangle\quad \text{for all $f\in C(K)$}.\] 
Here and below $\langle\cdot,\cdot\rangle$ denotes the dual pairing. Now let $\mathcal{F}_0 = \{f\in \mathcal{F}: f|_{V_0}=0\}$ and $\mathcal{F}^*$ (resp.\@ $\mathcal{F}_0^*$) be the dual of $\mathcal{F}$ (resp.\@ $\mathcal{F}_0$). 
A bounded function $\rho^H \in L^2(0,T, \mathcal{F})$ with $\partial_t \rho^H \in L^2(0,T,\mathcal{F}_0^*)$ is said to be a \emph{weak solution of (\ref{eq:hydrodynamiceqn})} if for every $t\in [0,T]$ and every $\varphi \in L^2(0,T,\mathcal{F}_0)$,
\begin{equation}\label{eqn:weaksol}
\int_0^t\int_K\, (\partial_s\rho^H_s) \varphi_s \,d\mu\:ds = -\int_0^t\, \mathcal{E}(\rho^H_s, \varphi_s)\,ds + \int_0^t\,\langle \chi(\rho^H_s)\partial H_s, \partial\varphi_s\rangle_{\mathcal{H}}\,ds,
\end{equation}
if $\rho^H_0=\rho_0$ in $L^2(K,\mu)$ and if $\rho_t^H-h\in\mathcal{F}_0$ for a.e. $t\in (0,T)$, where $h\in\mathcal{F}$ is the unique harmonic function on $K$ with boundary values $\bar\rho_i$ on $V_0$.
%\begin{equation}
%\langle u_T, F \rangle = \langle u_0, F \rangle + \int_0^T \, \left[\langle u_t, \Delta F\rangle - \int_K \, \chi(u_t) d\Gamma(H_t, F)\right]\,dt
%\end{equation}
%with Dirichlet boundary condition on $V_0$ and with initial condition $u_0$ satisfying $0\leq u_0\leq 1$. 
%{\color{blue} (Q by Michael: I think we have to test in time and space, at least if we want to apply Joe's / your uniqueness argument. See (5) in the uniqueness sketch. From point of view of existence of solution this is completely ok.)} 
The specification of the initial condition makes sense, because  under the required regularity conditions any such $\varrho^H$ will be an element of $C([0,T],L^2(K,\mu))$, see the references mentioned 
%in Section \ref{S:PDE} 
below. There we also comment on the existence and uniqueness of a weak solution $\varrho^H$ to (\ref{eq:hydrodynamiceqn}). 

%We can now state the law of.
\begin{theorem}[Joint density-current large numbers (LLN)]
\label{thm:LLN}
For every $t\in [0,T]$, $\delta>0$, $G \in C(K)$, and $F \in {\rm dom}\:\Delta$, we have
\[\lim_{N\to\infty} \mathbb{P}^N_{\eta^N}\left(\left|\langle \pi_t^N, G\rangle - \int_K \, G \rho^H_t \,d\mu\right|>\delta\right)=0\]
%and
\[\lim_{N\to\infty} \mathbb{P}^N_{\eta^N} \left(\left|\langle {\bf W}_t^N, \partial_N F\rangle
%_{L^2(E_N,\mu_N)} 
- \int_0^t \,\left\langle {\bf J}_{H_s}(\rho^H_s), \partial F\right\rangle_\mathcal{H} ds \right|>\delta\right)=0,\]
where $\rho^H$ denotes the unique weak solution of \eqref{eq:hydrodynamiceqn}, and for a.e. $s$ the vector field ${\bf J}_{H_s}(\rho_s^H)$ is defined by the identity
\[
\langle {\bf J}_{H_s}(\rho_s^H), \partial F\rangle_\mathcal{H}  := \mathcal{E}(\rho_s^H, F) + \langle \chi(\rho_s^H) \partial H_s, \partial F\rangle_\mathcal{H}
.\]
\end{theorem}
%{\color{blue}(Q by Michael: Please check.)}

%Now we state the main results pertaining to the hydrodynamic limit of the density-current pair $(\pi^N_t, {\bf W}^N_t)$.
%
%The LLN states that $(\pi^N_t, \partial_t{\bf W}^N_t)$ converges in $\mathbb{P}^N_{\mu^N,H}$-law to $(\rho_t^H, {\bf J}_{H_t}(\rho^H_t))$, where $\rho_t^H$ is the unique weak solution of (\ref{eq:hydrodynamiceqn}), and ${\bf J}_{H_t}(\rho_t^H)$ is defined via (\ref{eq:instcurrent}).

We also wish to quantify the probability of the (rare) event that a given trajectory deviates from the hydrodynamic solution \eqref{eq:hydrodynamiceqn}. This is done by proving a large deviations principle (LDP). Fix $\rho_0 \in C_e(K)$. 
Let 
\[\mathcal{FM}_0 := \{\rho(x)\,d\mu(x) : 0\leq \rho\leq 1~\mu\text{-a.e.},~\rho\in \mathcal{F}\}\] 
be the set of positive measures which are absolutely continuous with respect to $\mu$, whose density is bounded by $1$ and has finite energy.
We will work with $E:=D([0,T], \mathcal{FM}_0 \times \mathcal{H})$, the space of c\`adl\`ag paths from $[0,T]$ to $\mathcal{FM}_0\times \mathcal{H}$ endowed with the Skorokhod topology. 

%
%We fix an initial density profile $\rho_0: K \to\mathbb{R}$ which is bounded away from $0$ and from $1$, and has boundary value $\rho_0(a_i) = \bar\rho_i$. In order to single out the dynamical fluctuations about the hydrodynamic trajectory, we shall assume that the process $(\eta^N_t)_{t\in [0,T]}$ is started from a deterministic initial configuration $\eta^N_0 \in \mathcal{X}_N$, and that the sequence $\{\eta^N_0\}_N$ converges weakly to $\rho_0$: that is, for every $f\in C(K)$,
%\begin{align}
%\lim_{N\to\infty} \frac{1}{3^N} \sum_{x\in V_N} \eta^N_0(x) f(x) = \int_K \, \rho_0(x) f(x)\,\nu(dx).
%\end{align}
%
%Let $\Mzero$ be the family of nonnegative density functions with respect to $\nu$ on $K$, bounded above by $1$. Let $\Mone$ be the family of finite signed Borel measures on the space of $1$-forms $\Omega_1(K)$ on $K$. We denote by $C([0,T], \Mzero\times\Mone)$ the space of continuous $(\Mzero\times\Mone)$-valued paths in the time interval $[0,T]$, and by $D([0,T], \Mzero\times \Mone)$ the Skorokhod space of $(\Mzero\times \Mone)$-valued c\`{a}dl\`{a}g paths in the time interval $[0,T]$.
%\mathcal{FM}_0
For $H\in C^1([0,T], {\rm dom}\Delta)$, we introduce the functional $J_H:= J_{H,T,\rho_0} : E \to \mathbb{R}\cup \{+\infty\}$ as follows. If $(\pi, {\bf W}) \notin C([0,T], \mathcal{FM}_0\times \mathcal{H})$, set $J_H(\pi,{\bf W})=+\infty$; otherwise set
\begin{align*}
J_H(\pi, {\bf W}) &=  \langle \partial H_T, {\bf W}_T\rangle_\mathcal{H} - \int_0^T \, \left\langle \frac{\partial}{\partial t}\partial H_t, {\bf W}_t\right\rangle_\mathcal{H} \,dt - \int_0^T\, \langle \Delta H_t, \pi_t \rangle \,dt\\
\nonumber &  \quad -\int_0^T\,\left\langle \chi\left(\rho_t\right) \partial H_t, \partial H_t\right\rangle_\mathcal{H}\, dt ~+~ \sum_{i=0}^2 \bar\rho_i \int_0^T \, (\partial^\perp H_t)(a_i)\,dt.
\end{align*}
Here $\rho_t = \frac{d\pi_t}{d\mu}$ and $\partial^\perp$ denotes the normal derivative as defined in \cite{KigamiBook, StrichartzBook}. Put
\[
J(\pi, {\bf W}) = \sup_{H\in \domH} J_H(\pi, {\bf W}).\]

Let $\mathcal{A}$ be the set of all $(\pi, {\bf W}) \in C([0,T],\mathcal{FM}_0 \times \mathcal{H})$ satisfying $d\pi_0 = \rho_0 \, d\nu$, ${\bf W}_0=0$, and the \emph{conservation law} $\partial_t \pi_t + \partial^*(\dot{\bf W}_t)=0$ in the weak formulation: for every $\varphi \in \mathcal{F}_0$, $
\langle \varphi, \pi_t- \pi_0\rangle = \langle \partial\varphi, {\bf W}_t\rangle_\mathcal{H}.
$
We introduce the \emph{dynamical rate function}
\begin{align}
\label{eq:dynamicalratefcn}
I(\pi, {\bf W}) = \left\{\begin{array}{ll} J(\pi, {\bf W}), & \text{if}~(\pi, {\bf W})\in \mathcal{A},\\ +\infty,& \text{otherwise}. \end{array}\right.
\end{align}
It turns out that $J(\pi,{\bf W})$ can be written in the more symmetric form
\begin{align*}
J(\pi, {\bf W}) = \frac{1}{2}\int_0^T\, \left\langle[\chi(\rho_t)]^{-1} \left(\frac{\partial}{\partial t} {\bf W}_t + \partial \rho_t\right), \frac{\partial}{\partial t }{\bf W}_t + \partial \rho_t\right\rangle_\mathcal{H}\,dt.
\end{align*}

\begin{theorem}[Joint density-current LDP]
\label{thm:LDP}
For each closed set $\mathcal{C}$ and each open set $\mathcal{O}$ of $E$,
\begin{align*}
\limsup_{N\to\infty} \frac{1}{|V_N|} \log \mathbb{P}^N_{\eta^N}\left((\pi^N, {\bf W}^N)\in \mathcal{C} \right) &\leq -\inf_{(\pi, {\bf W})\in \mathcal{C}} I(\pi, {\bf W}),\\
 \liminf_{N\to\infty} \frac{1}{|V_N|} \log \mathbb{P}^N_{\eta^N}\left((\pi^N, {\bf W}^N)\in \mathcal{O} \right)&\geq -\inf_{(\pi, {\bf W})\in \mathcal{O}} I(\pi, {\bf W}).
\end{align*}
\end{theorem}

%\subsection

\bigskip\noindent{\bf Outline of proof strategy.}
Our proof strategy is conceptually aligned with the hydrodynamic limit program originating from \cite{KOV89,GPV88}, which has since been expounded in the monograph \cite{KipnisLandim} and applied to various low-dimensional lattice gas models. In particular we are influenced by the works \cite{BDGJL03, BDGJL07, BodineauLagouge}. For an earlier work on the hydrodynamic limit of a related particle system on $SG$ see \cite{Jara}.

That said, we had to overcome a number of technical obstacles to prove the limit theorems on $SG$. In a nutshell, the difficulties can be attributed to the well-known fact that on $SG$ (and other fractals), the energy measure is singular to the self-similar (Hausdorff) measure \cite{Kusuoka,BST99,Hino}. This has two consequences. On the microscopic level, there is no translational invariance, and one needs new tools (resistance-based energy inequalities) to establish a coarse-graining lemma in order to pass to the scaling limit. On the macroscopic level, one needs to utilize notions of vector calculus and (S)PDEs developed through Dirichlet forms.

For the sake of readability we have divided the proofs of Theorems \ref{thm:LLN} and \ref{thm:LDP} into several papers, whose contents are summarized as follows:

\emph{The moving particle lemma} \cite{ChenMPL}: we prove an energy inequality in the symmetric exclusion process on any finite weighted graph, bounding the cost of swapping particle configurations at vertices $x$ and $y$ in the exclusion process by the effective resistance distance $R_{\rm eff}(x,y)$ in the random walk process. The proof is based on the \emph{octopus inequality} of \cite{CLR09}, which was key to the positive resolution of Aldous' spectral gap conjecture.

\emph{A local ergodic (coarse graining) theorem} \cite{LocalErgodic}: we show that on many (strongly) recurrent graphs \cite{Telcs01, Telcs01_2,BCK05}, local functions of the occupation variables $\eta$ in the (suitably rescaled) exclusion process can be replaced by their macroscopic averages in the scaling limit, with probability superexponentially close to $1$. This local ergodic theorem is stated for the conservative version and the boundary-driven version of the exclusion process. It is proved using the aforementioned moving particle lemma, and can be applied to fractal graphs and trees that lack translational invariance.

\emph{Evolution equations on resistance spaces} \cite{EvolutionResistance}: we examine the solvability of evolution PDEs which generalize \eqref{eq:hydrodynamiceqn} on spaces which support Kigami's resistance forms \cite{Kigamiresistance}, using tools from Dirichlet forms and the induced vector analysis.

\emph{Hydrodynamic limit of the exclusion process on $SG$} \cite{SGHydro}: we utilize the results of the previous three papers, the vector analysis developed in \cite{HinzRocknerTeplyaev, HinzTeplyaevNS, HinzTeplyaevReview}, along with the established hydrodynamic limit program, to prove Theorems \ref{thm:LLN} and \ref{thm:LDP}.

%\subsection

\bigskip\noindent{\bf Semilinear evolution equations.}\label{S:PDE}
To verify the existence and uniqueness of a weak solution to problem (\ref{eq:hydrodynamiceqn}), we first consider the problem 
\begin{align}
\label{eqn:Dirichlet}
\left\{\begin{array}{lll} \partial_t w(t,x)&= \Delta w(t,x) - \partial^*\left(\chi(w(t,x)+h(x)) \partial H(t,x)\right) & \text{on}~(0,T)\times (K\setminus V_0), \\
 w(0,\cdot)&= \rho_0 -h,& \text{on}~K\setminus V_0,\\ 
 w(\cdot,a_i)&= 0& \text{on}~(0,T),\end{array}\right.
\end{align}
Here $h:K\to\mathbb{R}$ denotes the unique solution $h\in\mathcal{F}$ of the Dirichlet problem 
\begin{align}
\label{eqn:h}
\left\{\begin{array}{lll} \Delta h&=0 &\text{on}~K\setminus V_0,\notag\\
h(a_i)&=\bar\rho_i.\end{array}\right.\notag
\end{align}
We translate (\ref{eqn:Dirichlet}) into an abstract Cauchy problem and solve it using monotone operator methods \cite{Lions}. The (Dirichlet) Laplacian $\Delta$ may be viewed as a bounded variational operator $\Delta:\mathcal{F}_0\to\mathcal{F}_0^\ast$. We consider the nonlinear operator defined by
\[A(t,v):=-\Delta v+\partial^\ast(\chi(v+h)\partial H_t),\ \ v\in\mathcal{F}_0.\]
Recall that $H \in C^1([0,T],{\rm dom}\Delta) \subset L^\infty(0,T,\mathcal{F})$. Together with the resistance form properties of $(\mathcal{E},\mathcal{F})$ this can be used to see that writing
$(\mathcal{A}(u))_t:=A(t,u_t)$ for a given function $u:(0,T)\to\mathcal{F}_0$, we obtain a bounded and demicontinuous operator $\mathcal{A}:L^2(0,T,\mathcal{F}_0)\to L^2(0,T, \mathcal{F}_0^\ast)$ which satisfies 
\begin{equation}
\lim_{\left\|u\right\|_{L^2(0,T,\mathcal{F})\to\infty}} \frac{\left\langle \mathcal{A}(u),u\right\rangle}{\left\|u\right\|_{L^2(0,T,\mathcal{F})}}=+\infty.
\end{equation}
Here we use again $\left\langle \cdot,\cdot\right\rangle$ to denote the dual pairing in the obvious sense. We now rephrase (\ref{eqn:Dirichlet}) as the abstract Cauchy problem 
\begin{align}
\label{eqn:absCauchy}
\left\{\begin{array}{lll}
\partial_t w_t+A(t,w_t)&=0  &\text{for a.e. $t\in (0,T)$} \\
w(0)&=\rho_0-h.
\end{array}\right.
\end{align}
A function $w\in L^2(0,T, \mathcal{F}_0)$ with $\partial_t w\in L^2(0,T, \mathcal{F}_0^\ast)$ is called a \emph{(strong) solution} to the abstract Cauchy problem (\ref{eqn:absCauchy}) if the first identity holds in $\mathcal{F}^\ast_0$ (for a.e. $t\in (0,T)$) and the second holds in $L^2(K,\mu)$. Note that any $w\in L^2(0,T, \mathcal{F}_0)$ with $\partial_t w\in L^2(0,T, \mathcal{F}_0^\ast)$ is a member of $C([0,T],L^2(K,\mu))$, so that the second condition makes sense, see \cite[Chapter III, Proposition 1.2]{Showalter}. 

\begin{proposition}
There exists a solution $w$ to the abstract Cauchy problem (\ref{eqn:absCauchy}).
\end{proposition}
This follows from \cite[Th\'eor\`eme 2.1]{Lions} together with certain estimates based on the resistance form properties of $(\mathcal{E},\mathcal{F})$. Given a solution $w$ of (\ref{eqn:absCauchy}), the function 
$\rho^H:=w+h$
is an element of $L^2(0,T,\mathcal{F})$ and satisfies $\partial_t\rho^H\in L^2(0,T,\mathcal{F}^\ast_0)$. We also have
\[\partial_t \rho^H_t= \Delta \rho^H_t - \partial^*\left(\chi(\rho^H_t) \partial H_t\right) \quad\text{for a.e. $t\in (0,T)$}, \]
seen in $\mathcal{F}_0^\ast$, and this implies the validity of the integral identity (\ref{eqn:weaksol}).
Moreover, we observe that $\rho^H_0=\rho_0$ and $\rho^H(t,a_i)=\bar\rho_i$ for a.e. $t\in (0,T)$.

\begin{theorem}
The function $\rho^H$ is the unique weak solution to (\ref{eq:hydrodynamiceqn}).
\end{theorem}
The uniqueness can be obtained by showing that the difference of two solutions has $L^1(K,\mu)$-norm decreasing in time. This argument was already used in \cite[Appendix]{BodineauLagouge} and \cite[Proof of Lemma 7.1]{Landimetal}. It relies on the Markov property of $(\mathcal{E},\mathcal{F})$. Details are provided in \cite{EvolutionResistance}.

%\subsection

\bigskip\noindent{\bf Perspectives.}
Theorem \ref{thm:LLN} may be viewed as a first-principles derivation of a fluid equation on a fractal, subject to external biases on the boundary. The solution of this equation gives the expected trajectory of the macroscopic density and current. Theorem \ref{thm:LDP} characterizes the fluctuations about this expected trajectory in terms of the large deviations rate function $I(\pi, {\bf W})$. 

Let us note that on simple graphs which either are discrete tori or have two boundary points (such as $\mathbb{Z} \cap [-n, n]$), there is an alternative combinatorial approach to deriving the rate function \cite{DerridaICM}. However it is unclear if this approach generalizes to $SG$ (with the standard $3$-point boundary) or other infinite weighted graphs. This explains why we followed the hydrodynamic limit program, which has a more robust analytic flavor.

The rate function can be used to analyze properties of macroscopic fluctuations in boundary-driven diffusive processes on networks, which is a current topic of interest in nonequilibrium statistical mechanics; see the excellent recent review \cite{BDGJL15}. Some highlights in this area include the universality of the cumulant of the mean long-time current in networks with two boundary points \cite{ABDS13}, under the hypothesis of the \emph{additivity principle} \cite{BDAdditivity}. It is also of interest to investigate the validity of the additivity principle in closely related boundary-driven particle systems \cite{BDGJL07}. We hope to investigate these and related problems on $SG$ (and other fractals) in the near future, but keep in mind that there are key differences that make the analysis more difficult than on tori and simple lattices. We list a few open related questions:
\begin{itemize}
	\item Analytically or numerically characterize the infimum of the rate function $J(\pi,{\bf W})$ as explicitly as possible.
	\item Describe the motion of a tagged particle in the exclusion process on $SG$.
	\item How does the aforementioned current cumulant problem manifest on $SG$?
	\item Investigate asymmetric exclusion processes and other non-gradient-type particle systems on fractals and other weighted graphs. A key technical step would be to identify energy (spectral gap) inequalities relevant to each process or, in some cases, make use of extra regularity following 
	\cite{kajino,kelleher,baudoin}.\end{itemize}

\bibliographystyle{spmpsci}
\bibliography{biblio_hydro}
\end{document}